\documentclass[conference]{IEEEtran}
\IEEEoverridecommandlockouts

\usepackage{cite}
\usepackage{amsmath,amssymb,amsfonts}
\usepackage{algorithmic}
\usepackage{graphicx}
\usepackage{textcomp}
\usepackage{multirow} 
\usepackage{float} 
\usepackage{xcolor}
\usepackage{hyperref}
\hypersetup{
    colorlinks=true,
    linkcolor=black,
    citecolor=blue,
    urlcolor=blue,
}
\def\BibTeX{{\rm B\kern-.05em{\sc i\kern-.025em b}\kern-.08em
    T\kern-.1667em\lower.7ex\hbox{E}\kern-.125emX}}

\title{Sequence Spreading-Based Semantic Communication Under High RF Interference}

\author{
    Hazem Barka\IEEEauthorrefmark{1}, 
    Georges Kaddoum\IEEEauthorrefmark{1}, 
    Mehdi Bennis\IEEEauthorrefmark{2}, 
    Md Sahabul Alam\IEEEauthorrefmark{3}, 
    Minh Au\IEEEauthorrefmark{4}\\
    \IEEEauthorblockA{
        \IEEEauthorrefmark{1}Electrical Engineering Department, ETS, Montreal, QC\hspace{2pt} H3C 1K3, Canada\\ 
    Email: hazem.barka.1@ens.etsmtl.ca, georges.kaddoum@etsmtl.ca
    }
    \IEEEauthorblockA{
        \IEEEauthorrefmark{2} Centre for Wireless Communications, University of
Oulu, Oulu\hspace{2pt} 90014, Finland
\\ 
        Email: mehdi.bennis@oulu.fi
    }
    \IEEEauthorblockA{
        \IEEEauthorrefmark{3} Electrical and Computer Engineering Department, California State University, Northridge, CA\hspace{2pt} 91330, United States\\ 
        Email: md-sahabul.alam@csun.edu
    }
    \IEEEauthorblockA{
        \IEEEauthorrefmark{4}Hydro-Québec Research Institute (IREQ),  Varennes, QC\hspace{2pt} J3X 1S1, Canada\\ 
        Email: au.minh2@hydroquebec.com
    }
}

\begin{document}

\maketitle

\begin{abstract}
   In the evolving landscape of wireless communications, semantic communication (SemCom) has recently emerged as a 6G enabler that prioritizes the transmission of meaning and contextual relevance over conventional bit-centric metrics. However, the deployment of SemCom systems in industrial settings presents considerable challenges, such as high radio frequency interference (RFI), that can adversely affect system performance. To address this problem, in this work, we propose a novel approach based on integrating sequence spreading techniques with SemCom to enhance system robustness against such adverse conditions and enable scalable multi-user (MU) SemCom. In addition, we propose a novel signal refining network (SRN) to refine the received signal after despreading and equalization. The proposed network eliminates the need for computationally intensive end-to-end (E2E) training while improving performance metrics, achieving a 25\% gain in BLEU score and a 12\% increase in semantic similarity compared to E2E training using the same bandwidth.
\end{abstract}

\section{Introduction}
\subsection{Motivations}
In industrial settings, communication systems are nowadays increasingly tasked with handling complex, dynamic, and mission-critical operations. However, conventional communication methods, which transmit data with a strict focus on bit-level accuracy, frequently prove to be inefficient in these scenarios. The massive amount of redundant information can quickly overwhelm the network, leading to congestion and inefficiencies. In this context, by prioritizing only meaningful or relevant data transmission, semantic communication (SemCom) offers a promising solution to improve spectral efficiency \cite{deepsc_high_RFI,deepsc,deepsc_r,deespc_s}. This is particularly crucial in high-density industrial environments where available bandwidth is limited.

However, while SemCom holds promise for industrial applications due to its ability to prioritize essential data, it remains vulnerable without adequate interference mitigation strategies. High radio frequency interference (RFI) scenarios are prevalent in industrial plants, power grids, and near heavy machinery, where it is generated by high-voltage equipment, motors, and certain wireless network configurations \cite{measurements,bnet,mitigation_techniques,markov_middleton_model,ours_LLR}. Such an abundance of impulsive and continuous interference sources leads to unpredictable communication disruptions. Thus, conventional communication systems, which rely on precise bit-level transmission, experience difficulties with preserving signal fidelity amidst strong interference \cite{measurements,bnet,mitigation_techniques,markov_middleton_model,ours_LLR}. While focusing on semantic aspects can inherently make SemCom systems more resilient to certain types of noise, particularly minor bit errors, high RFI conditions can still pose significant challenges \cite{deepsc_high_RFI}. Interference not only corrupts the transmitted signal but can also distort the semantic meaning of the data, leading to communication errors that go beyond simple bit-level issues \cite{deepsc_high_RFI}. Accordingly, there is a critical need to develop techniques capable of enhancing the robustness of SemCom systems in such environments \cite{deepsc_high_RFI}.
\subsection{Related Work}
In recent years, several single-user (SU) deep SemCom systems were developed, with the most well-known among them being DeepSC \cite{deepsc}. DeepSC is a text-based deep learning (DL) system designed to maximize transmission capacity while minimizing semantic errors between the transmitted and received text. Variants of DeepSC were proposed for different tasks, including speech transmission \cite{deespc_s}, image communication \cite{image_VQ}, and context-based applications \cite{cb_sc}. These systems emphasize semantic accuracy over bit-level precision, providing more meaningful communication under noise and fading. However, previous studies primarily focused on SU scenarios and did not address scalability and robustness challenges encountered with multi-user (MU) or high interference levels.

For MU SemCom scenarios, power-domain non-orthogonal multiple access (PD-NOMA) was the preferred multiple access method, as several studies used this approach \cite{NOMA_two_users_joint_training_1,NOMA_two_users_joint_training_2,NOMA_two_users_joint_training_3,NOMA_two_users_joint_training_4}. However, despite their effectiveness, PD-NOMA-based solutions often require extensive joint end-to-end (E2E) online training and were found to have limitations in scalability, typically supporting two or three users on a shared channel without significant performance degradation \cite{NOMA_two_users_joint_training_1,NOMA_two_users_joint_training_2,NOMA_two_users_joint_training_3,NOMA_two_users_joint_training_4}. Other conventional multi-access methods, such as orthogonal frequency division multiple access (OFDMA), code division multiple access (CDMA), and code-domain NOMA (CD-NOMA), remain largely unexplored within the SemCom framework. 

Moreover, in \cite{deepsc_high_RFI}, the authors argued that strong RFI can significantly degrade SemCom's performance, potentially destroying the faithfulness of transmitted information. Accordingly, the authors introduced a probabilistic framework to characterize the performance limits of DeepSC \cite{deepsc} and validated them through extensive simulations, highlighting the need for robust SemCom systems in the face of strong interference.

Yet, despite these advances, none of the previous studies has specifically addressed the development of techniques to enhance the resilience of SemCom systems against high RFI. Accordingly, the unique challenges posed by severe interference levels in such settings remain critical areas requiring further research \cite{deepsc_high_RFI}. Developing robust interference mitigation strategies is essential to facilitate the practical deployment of SemCom systems in industrial and other high RFI scenarios, thus ensuring reliable communication, particularly in scenarios where conventional methods struggle.

\subsection{Contributions}

In this study, seeking to address the challenges posed by high RFI in SemCom systems, we propose a novel solution to enhance their robustness and scalability by:

\begin{itemize}
    \item introducing an AI-assisted sequence spreading (SS) approach tailored for SemCom systems, which is robust against high RFI (typically observed in industrial communications) and enables scalable MU capabilities;
    \item proposing a novel signal refining network (SRN) to refine the received signal post-despreading and equalization. The proposed network significantly enhances system performance by using the interference itself as an aid rather than a detriment. In addition, the SRN eliminates the need for computationally intensive online E2E training.
\end{itemize}

\section{System Model}
\subsection{Signal Model}
\begin{figure*}[t]
\centering
\includegraphics[width=1\linewidth]{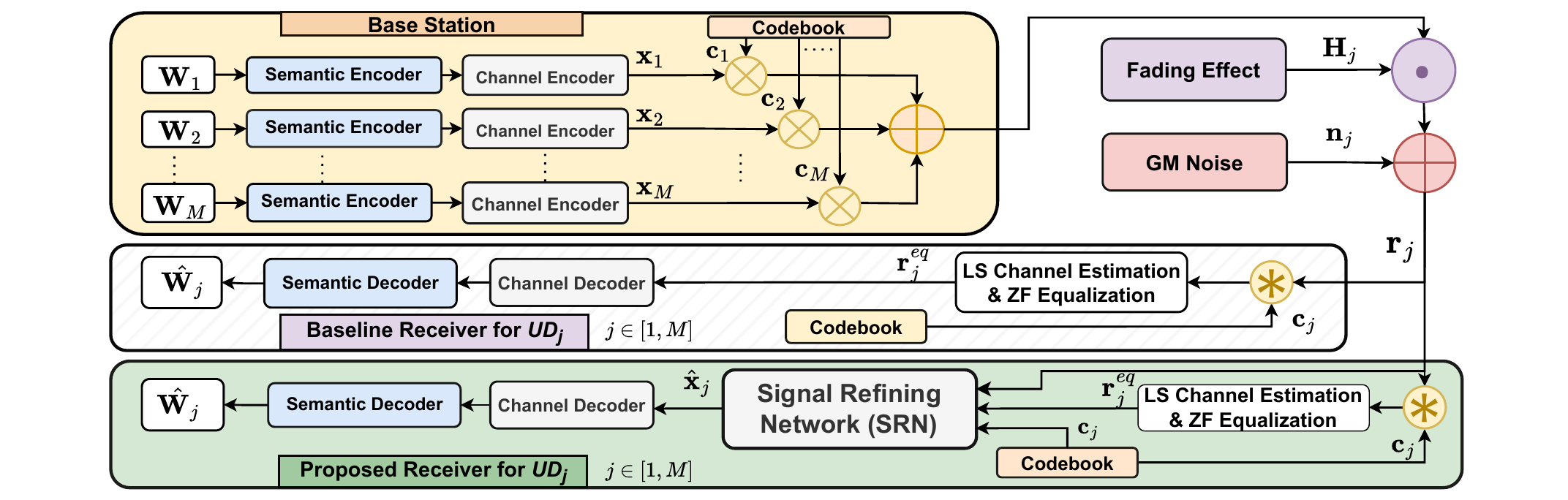}
\caption{The considered SS-based deep SemCom system model with $M$ users.} \label{fig:system_model}
\end{figure*}

We consider a downlink text-based SemCom system model, illustrated in Fig. \ref{fig:system_model}, where multiple user devices  (\( UD \)s) receive signals through a shared channel. The received signal vector at user device \( UD_j \) in a time-varying flat-fading channel is expressed as shown in Eq. \eqref{eq:rj}:
\begin{equation}
    \mathbf{r}_j = \sum_{i=1}^{M} \mathbf{H}_j(\mathbf{x}_i\otimes\mathbf{c}_i) + \mathbf{n}_j,
    \label{eq:rj}
\end{equation}
where
\begin{itemize}

    \item \( M \): number of user devices in the network.
    \item \(\mathbf{H}_j \): diagonal Rayleigh fading channel matrix for  \( UD_j \). 
    \item \( \mathbf{c}_i \): SS code vector for \( UD_i \).
    \item \( \mathbf{x}_i \): vector of transmitted symbols for \( UD_i \).
    \item \( \mathbf{n}_j \): external noise and RFI vector at user \( UD_j \).
    \item  \( \otimes \): the Kronecker product between 2 vectors.
\end{itemize}

It is important to note the following: \( \mathbf{x}_i \in \mathbb{C}^K \), with \( K \) representing the number of transmitted symbols for a given sequence of words (denoted by \( \mathbf{W}_i \)); \( \mathbf{H}_j \in \mathbb{C}^{K\times K}\); \( \mathbf{c}_i \in \mathbb{R}^{SF} \), with \( SF \) being the spreading factor; and \( \mathbf{r}_j \) and \( \mathbf{n}_j \in \mathbb{C}^{K \times SF} \).

\subsection{Noise and RFI Model}
External noise and RFI vector \( \mathbf{n}_j \) at \( UD_j \) is assumed to be dominated by high RFI, which is modeled as a Gaussian mixture (GM) with \( S \) distinct states. Each state of the GM model represents a different interference level, characterized by its mean and variance. The noise probability density function for \( \mathbf{n}_j[k] \), where \( k \in \{1,..,K\}\), is given by Eq. \eqref{eq:nj}:
\begin{equation}
   \mathbf{n}_j[k] \sim \sum_{s=1}^{S} \pi_s \mathcal{N}(\mu_s, \sigma_s^2),
   \label{eq:nj}
\end{equation}
where \( \pi_s \) is the weight (probability) of the \( s \)-th Gaussian component, with \( \sum_{s=1}^{S} \pi_s = 1 \). Furthermore, \( \mu_s \) and  \( \sigma_s^2 \) are the mean and the variance of the \( s \)-th Gaussian component, respectively. In particular, the variance of the first Gaussian component, \( \sigma_1^2 \), is related to the thermal noise power spectral density \( N_0 \), where \( \sigma_1^2 = N_0 B \), and \( B \) represents system bandwidth.
 This GM model effectively captures the impulsive nature of high RFI environments where the interference can considerably fluctuate between different levels of severity, leading to varying noise characteristics across different time instances and components of the received signal \cite{measurements}.
\subsection{Despreading, Channel Estimation, and Equalization}
\label{subsection:equalize}
In this subsection, we detail the despreading process, channel equalization, and estimation. Of note, equalization is performed after despreading the received signal. First, user \( UD_j \) despreads the received signal by correlating it with \( \mathbf{c}_j \) using convolution operation with stride $SF$, denoted by $\circledast$:
\begin{align}
\mathbf{r}_j^{\text{ds}} &= \mathbf{r}_j \circledast \mathbf{c}_j \\
&= \left( \sum_{i=1}^{M} \mathbf{H}_j \left( \mathbf{x}_i \otimes \mathbf{c}_i \right) + \mathbf{n}_j \right) \circledast \mathbf{c}_j \\
&= \sum_{i=1}^{M}\mathbf{H}_j \left( \mathbf{x}_i \otimes  \mathbf{c}_i \circledast \mathbf{c}_j \right)  + \mathbf{n}_j \circledast \mathbf{c}_j 
\label{eq:separated_signal}.
\end{align}

After despreading, we estimate channel coefficients $\hat{\mathbf{H}}_j$ using pilot symbols and the least squares (LS) method. Then, we apply zero-forcing (ZF) equalization to compensate for the channel effects. The equalized signal is calculated as shown in Eq \eqref{eq:equalized_signal}:
\begin{align}
\mathbf{r}_j^{\text{eq}} &=\underbrace{\hat{\mathbf{H}}^{-1}_j\mathbf{H}_j \left( \mathbf{x}_j \otimes \mathbf{c}_j \circledast \mathbf{c}_j \right)}_{\text{Desired signal}}+ \underbrace{\hat{\mathbf{H}}^{-1}_j(\mathbf{n}_j \circledast \mathbf{c}_j)}_{\text{Noise and RFI}} \nonumber\\ &+ \underbrace{\sum_{\substack{i=1 \\ i \ne j}}^{M}\hat{\mathbf{H}}^{-1}_j\mathbf{H}_j\left( \mathbf{x}_i \otimes \mathbf{c}_i \circledast \mathbf{c}_j \right)}_{\text{Inter-user interference}}, \label{eq:equalized_signal}
\end{align}
with $\hat{\mathbf{H}}^{-1}_j$ is the inverse of $\hat{\mathbf{H}}_j$. We observe that the noise and RFI term in Eq. \eqref{eq:equalized_signal} significantly impacts system performance. When the estimated channel coefficient $\hat{\mathbf{H}}_{j}[k]$ has a low magnitude---e.g., during deep fades---the division by $\hat{\mathbf{H}}_{j}[k]$ amplifies the noise and RFI components, thereby reducing the accuracy of semantic decoding. Moreover, high RFI environments not only increase the magnitude of the noise and RFI term but also cause errors in channel estimation, leading to improper equalization. These inaccuracies exacerbate the amplification of noise and RFI, further degrading performance. However, the properties of the spreading code $\mathbf{c}_j$ can help mitigate the impact of the noise and RFI term. 

\subsection{DL-Based Semantic and Channel Encoding/Decoding}

The proposed SemCom system uses a conventional offline-trained DeepSC model \cite{deepsc } for semantic and channel encoding at the transmitter and decoding at the receiver. At the transmitter, a semantic encoder \( S_{\hat{\boldsymbol{\beta}}}(\cdot) \) and a channel encoder \( C_{\hat{\boldsymbol{\alpha}}}(\cdot) \) map an $L$-word \( \mathbf{W}_j = [w_1,w_2,…,w_L]\) into a lower-dimensional space for transmission $\mathbf{x}_j \in \mathbb{C}^K$. At the receiver, after despreading and equalization (as detailed in Section \ref{subsection:equalize}), \( UD_j \) obtains the equalized signal \( \mathbf{r}_j^{\text{eq}} \). This signal is then processed by a channel decoder \( C_{\hat{\boldsymbol{\delta}}}(\cdot) \) and a semantic decoder \( S_{\hat{\boldsymbol{\theta}}}(\cdot) \) to reconstruct the sequence \( \hat{\mathbf{W}}_j \) (see Eq. \eqref{eq:xj}-\eqref{eq:wj}):
\begin{align} \mathbf{x}_j&= C_{\hat {\boldsymbol {\alpha }}}(S_{\hat {\boldsymbol {\beta }}}(\mathbf{W}_j)),\label{eq:xj}\\
 \hat {\mathbf{W}_j}&= S_{\hat {\boldsymbol {\theta }}}(C_{\hat {\boldsymbol {\delta }}}(\mathbf{r}^{\text{eq}}_j)).\label{eq:wj}
\end{align}

When trained in an E2E manner (see \cite{deepsc}) on fading channels and GM noise and RFI, the DeepSC architecture demonstrates suboptimal performance. This outcome is primarily due to the complex nature of the physical channel and the interference encountered (see Section IV). To address these challenges, we propose a novel alternative approach. Initially, we train the DeepSC model offline using a flat Rayleigh fading channel and AWGN. This pre-training stabilizes the semantic encoder and decoder, as well as the channel encoder and decoder. 

\section{SRN-assisted Sequence Spreading}
 \begin{figure*}[t]
\centering
\includegraphics[width=0.7\linewidth]{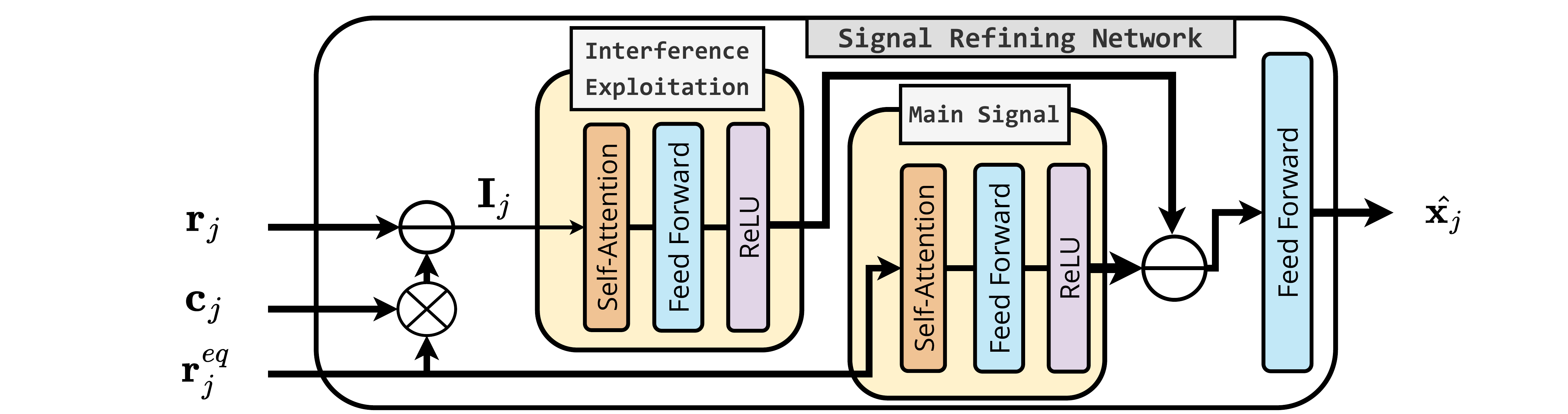}
\caption{The proposed SRN for SS-based Deep SemCom systems.} \label{fig:SRN}
\end{figure*}
\subsection{Sequence Spreading Codes}

To enhance the robustness of SemCom systems against high RFI, we propose using SS. This method involves spreading the signal \( \mathbf{x}_j \) of each $UD_j$ over a wider bandwidth $B$ using a distinct spreading code \( \mathbf{c}_j \), thereby increasing the signal's resistance to narrowband noise and RFI by distributing its power across a broader spectral range.

Numerous coding techniques for spread spectrum systems have been proposed, including m-sequences, Gold codes, Kasami sequences, and Walsh codes. Each technique offers advantages in terms of autocorrelation and cross-correlation properties. However, in this study, we focus on pseudo-random codes with values \( \{ -1, 0, 1 \} \). These codes exhibit low cross-correlation properties and feature a variable density parameter \( D \), which represents the proportion of non-zero elements in the sequence. Another critical parameter is the length of these spreading codes, known as the spread factor ($SF$). While a higher $SF$ increases the robustness of the signal against narrowband interference, this is achieved at the cost of increased bandwidth, as more symbols are required to transmit the same amount of information. Therefore, selecting an optimal $SF$ is essential to balance system performance in terms of throughput and interference resistance.

The spread sequence for each \( UD_j \) is generated and normalized according to the following steps:

\begin{itemize}
    \item \textbf{Sequence generation:}
    Each element of the spreading code \( \mathbf{c}_j \) is generated independently according to a random process with the following probabilities:
    \[
    \mathbf{c}_j[m] = 
    \begin{cases} 
    1 & \text{with probability } \frac{D}{2}, \\
    -1 & \text{with probability } \frac{D}{2}, \\
    0 & \text{with probability } 1-D,
    \end{cases}
    \]
    with \( m \) indexes the sequence elements.
    \item \textbf{Normalization:}
    To ensure power conservation, the sequence \( \mathbf{c}_j \) is normalized such that the autocorrelation at zero lag equals one ($\mathbf{c}_j \circledast \mathbf{c}_j = \mathbf{c}_j \cdot \mathbf{c}_j^T=1$, with $\mathbf{c}_j^T$ is the transpose of $\mathbf{c}_j$). This is achieved as follows:
    \begin{equation}
    \mathbf{c}_j \leftarrow \frac{\mathbf{c}_j}{\sqrt{\sum_{m=1}^{SF} |\mathbf{c}_j[m]|^2}}
    \end{equation}
\end{itemize}

\subsection{Signal Refining Network (SRN)}

The SRN is specifically designed to enhance the performance of SS-based SemCom systems operating under high RFI conditions. The SRN focuses on refining the received signal to better estimate the original transmitted signal vector \( \mathbf{x}_j \), leveraging the received signal vector \( \mathbf{r}_j \), the equalized despread vector \( \mathbf{r}_j^{\text{eq}} \), and the spreading code \( \mathbf{c}_j \).

The SRN initiates its process by re-spreading the equalized despread vector \( \mathbf{r}_j^{\text{eq}} \) and subtracting it from the original received signal vector \( \mathbf{r}_j \). This subtraction results in a residual interference vector \( \mathbf{I}_j \) as shown in Eq. \eqref{eq:Ij}:

\begin{equation}
\mathbf{I}_j = \mathbf{r}_j- (\mathbf{r}_j^{\text{eq}} \otimes \mathbf{c}_j).\label{eq:Ij}
\end{equation}

Traditionally, in SS-based systems, \( \mathbf{I}_j \) is discarded as it does not carry any useful data for  \( UD_j \). However, our approach recognizes the potential of \( \mathbf{I}_j \) to provide valuable insights into fading, noise, and interference patterns. In the present study, these frequently discarded elements are exploited through a neural network module. The residual interference vector \( \mathbf{I}_j \) undergoes a feature extraction process, where patterns of interference are analyzed and exploited. This information is integrated with data extracted from \( \mathbf{r}_j^{\text{eq}} \), thus enabling a more accurate reconstruction of the original transmitted symbol \( \mathbf{x}_j \). Consequently, the SRN significantly enhances system resilience to the environmental and electronic noise sources prevalent in high RFI environments.

Self-attention mechanisms in the SRN are used to analyze the inputs ( \( \mathbf{r}_j^{\text{eq}} \) and \( \mathbf{I}_j \)) and to dynamically prioritize segments of the signal that are most strongly affected by fading, noise, and RFI. This capability ensures that instead of treating all data uniformly, the network focuses on the parts that are crucial for reconstructing the original transmitted vector \(\mathbf{x}_j \). 

The SRN is trained to minimize the mean squared error (MSE) between the estimated (\( \hat{\mathbf{x}_j} \)) and original (\( \mathbf{x}_j \)) vectors:
\begin{equation}
    \text{MSE} = \frac{1}{K} \sum_{k=1}^{K} |\hat{\mathbf{x}}_{j}[k] - \mathbf{x}_{j}[k]|^2
\end{equation}

\begin{table}[t]
\caption{\label{tab:training_details}The Experimental Setup.}
\resizebox{0.9\columnwidth}{!}{
\renewcommand{\arraystretch}{1.05}
\begin{tabular}{|cc|c|}
\hline
\multicolumn{2}{|c|}{Parameter}                                                                                                                                                                                           & Value                                                         \\ \hline
\multicolumn{1}{|c|}{\multirow{5}{*}{\begin{tabular}[c]{@{}c@{}}Semantic \\ encoder/decoder\\ parameters\end{tabular}}} & Training mode                                                                                   & Offline                                                       \\ \cline{2-3} 
\multicolumn{1}{|c|}{}                                                                                                  & \begin{tabular}[c]{@{}c@{}}Number of transformer\\ layers\end{tabular}                          & 3                                                        \\ \cline{2-3} 
\multicolumn{1}{|c|}{}                                                                                                  & \begin{tabular}[c]{@{}c@{}}Embedding dimension for\\ the transformer layers\end{tabular}        & 256                                                           \\ \cline{2-3} 
\multicolumn{1}{|c|}{}                                                                                                  & \begin{tabular}[c]{@{}c@{}}Number of heads for \\ the transformer layers\end{tabular}           & 8                                                             \\ \cline{2-3} 
\multicolumn{1}{|c|}{}                                                                                                  & Sentence length                                                                                 & 30 words                                                      \\ \hline
\multicolumn{1}{|c|}{\multirow{3}{*}{\begin{tabular}[c]{@{}c@{}}Channel \\ encoder/decoder\\ parameters\end{tabular}}}  & Training mode                                                                                   & Offline                                                       \\ \cline{2-3} 
\multicolumn{1}{|c|}{}                                                                                                  & \begin{tabular}[c]{@{}c@{}}Number of feed\\ forward layers\end{tabular}                         & 2                                                             \\ \cline{2-3} 
\multicolumn{1}{|c|}{}                                                                                                  & \begin{tabular}[c]{@{}c@{}}Number of neurons\\ for each layer\end{tabular}                      & 256  
                                          \\ \cline{2-3} 
\multicolumn{1}{|c|}{}                                                                                                  & \begin{tabular}[c]{@{}c@{}}Number of symbols\\ for each word\end{tabular}                      & 8     \\ \hline
\multicolumn{1}{|c|}{\multirow{2}{*}{SRN parameters}}                                                                   & Training mode                                                                                   & Online                                                        \\ \cline{2-3} 
\multicolumn{1}{|c|}{}                                                                                                  & \begin{tabular}[c]{@{}c@{}}Number of neurons for\\ each feed forward layer\end{tabular}          & 256                                                           \\ \hline
\multicolumn{1}{|c|}{\multirow{3}{*}{\begin{tabular}[c]{@{}c@{}}Training\\ parameters\end{tabular}}}                    & Optimizer                                                                                       & Adam                                                          \\ \cline{2-3} 
\multicolumn{1}{|c|}{}                                                                                                  & Learning rate                                                                                   & 0.001                                                         \\ \cline{2-3} 
\multicolumn{1}{|c|}{}                                                                                                  & \begin{tabular}[c]{@{}c@{}}Number of training\\ epochs\end{tabular}                             & 20                                                            \\ \hline
\multicolumn{1}{|c|}{\multirow{6}{*}{\begin{tabular}[c]{@{}c@{}}Channel and \\ interference\\ parameters\end{tabular}}} & Coherence time                                                                                  & 16 symbols \\ 
\cline{2-3} 
\multicolumn{1}{|c|}{}                                                                                                  & \% of pilot symbols                                                                             & 25\%                                                          \\ \cline{2-3} 
\multicolumn{1}{|c|}{}                                                                                                  & \begin{tabular}[c]{@{}c@{}}Number of noise\\ states \( S \)\end{tabular}                        & 3                                                             \\ \cline{2-3} 
\multicolumn{1}{|c|}{}                                                                                                  & \begin{tabular}[c]{@{}c@{}}GM weights \\ \(\pi_s \forall s\)\end{tabular}                       & {[}0.9, 0.05, 0.05{]}                                         \\ \cline{2-3} 
\multicolumn{1}{|c|}{}                                                                                                  & \begin{tabular}[c]{@{}c@{}}GM power coefficients\\ \(\sigma_s^2/B\) \(\forall s\)\end{tabular} & \(\begin{bmatrix}
SF \cdot N_0 \\
30\cdot N_0   \\
50\cdot N_0   \\
\end{bmatrix}\)                                              \\ \hline
\end{tabular}}
\end{table}

\section{Performance Evaluation}
\begin{figure}[t]
    \centering
    \includegraphics[width=0.91\columnwidth]{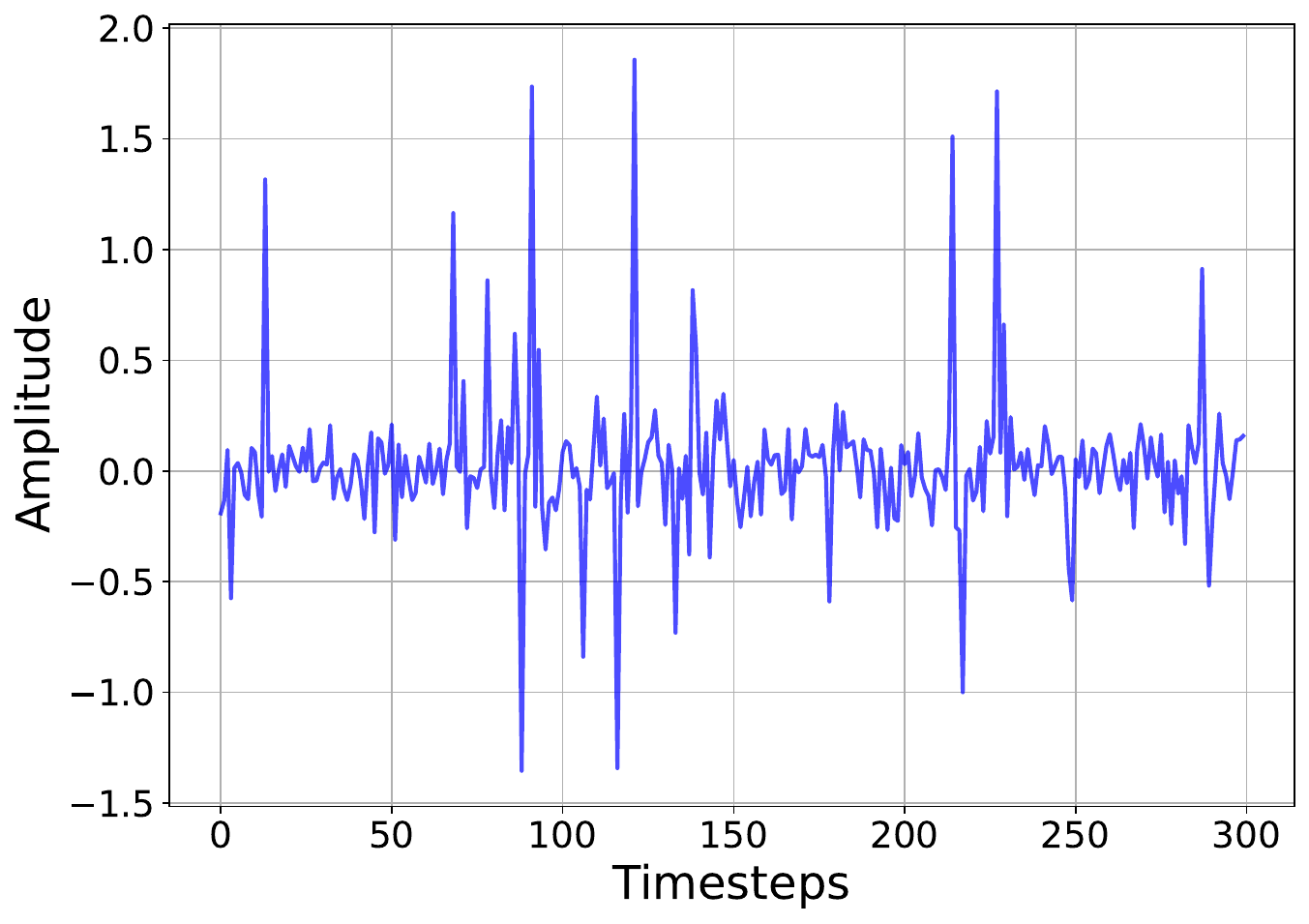}
\caption{Visualization of the GM-modeled noise and high RFI. The graph illustrates the impulsive and fluctuating nature of RFI commonly observed in industrial environments.}
    \label{fig:high_RFI_GM}
\end{figure}

In this section, we evaluate the performance of the proposed SRN-assisted SS-based SemCom system through simulations under various conditions. Specifically, we analyze the effects of the spreading factor (\( SF \)) and the signal-to-noise ratio (SNR = \( E_s/N_0 \), where \( E_s \) represents the average energy per symbol of \( \mathbf{x}_j[k] \)) in SU scenarios. To evaluate system scalability and robustness, this analysis is extended to MU scenarios. For training and testing in our experiments, we use the European Parliament text dataset \cite{europarl}. The training loss function and performance metrics, including BLEU score and semantic similarity, were previously thoroughly described in \cite{deepsc}. A sample of the considered GM-modeled noise and RFI variable $\mathbf{n}_j[k]$ is illustrated in Fig. \ref{fig:high_RFI_GM}. Further details on our experimental setup are provided in Table \ref{tab:training_details}.

\subsection{Effect of Spreading Factor \( SF \) in SU Scenarios}
\label{subsection:SF_effect}

\begin{figure}[t]
    \centering
    \includegraphics[width=0.96\columnwidth]{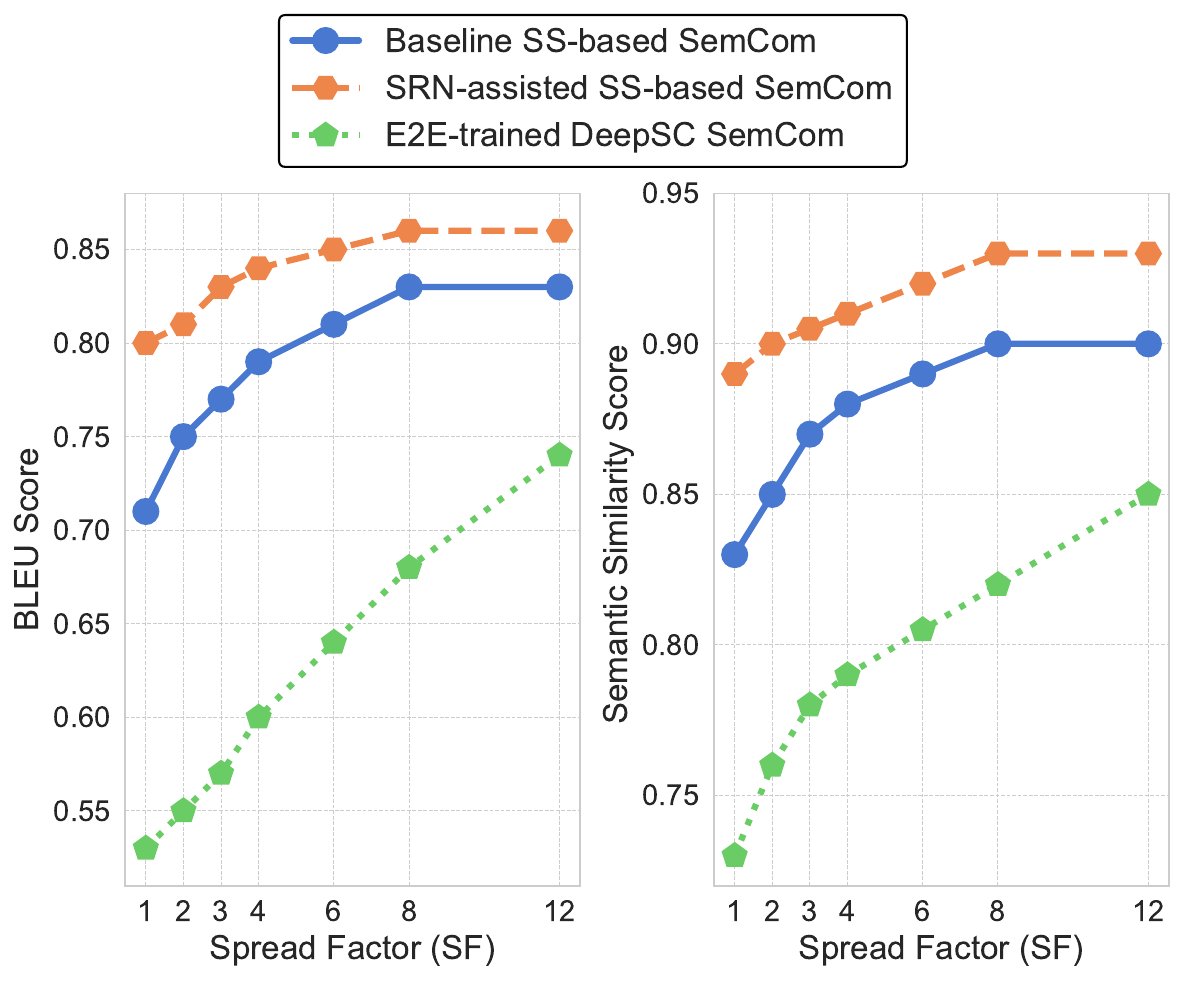}
    \caption{Performance comparison of baseline SS-based SemCom, SRN-assisted SS-based SemCom, and E2E-trained DeepSC SemCom systems across different $SF$ values with \( D = 1 \) and \( E_s/N_0 = 15 \) dB under Rayleigh fading and high RFI channel. For E2E-trained DeepSC, we increase the number of transmitted symbols by a factor of \( SF \).}
    \label{fig:BLEU_SemSim_vs_SF}
\end{figure}

\begin{figure}[t]
    \centering
    \includegraphics[width=1\columnwidth]{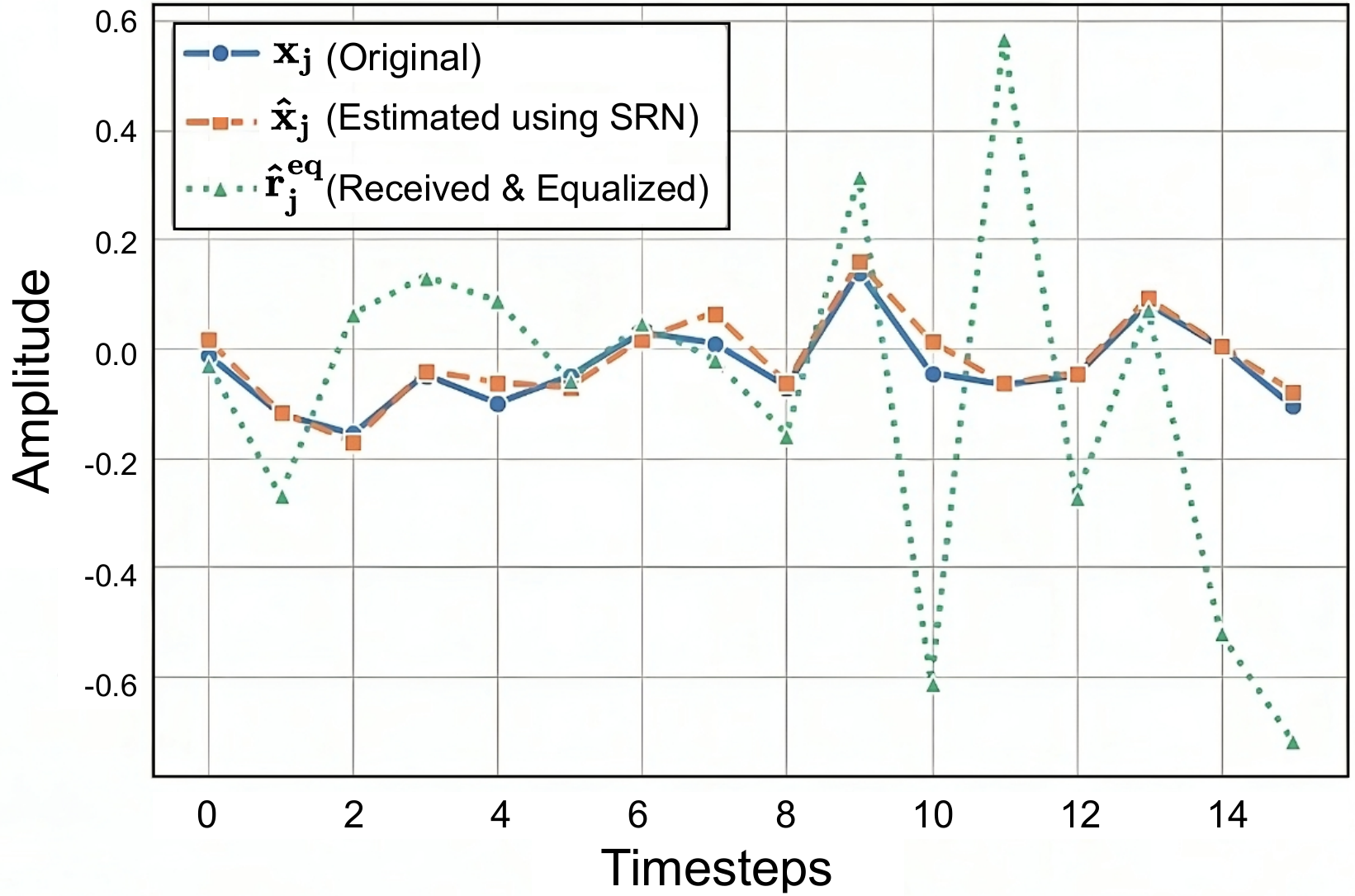}
    \caption{Illustration of SRN's signal estimation quality (the real part) in the SU scenario with \( E_s/N_0 = 15 \) dB and $SF = 1$.}
    \label{fig:signal_estimation}
\end{figure}
In the experiments illustrated in Fig. \ref{fig:BLEU_SemSim_vs_SF}, we examine the impact of $SF$ on system performance in SU scenarios under Rayleigh fading and high RFI conditions. As shown in Fig. \ref{fig:BLEU_SemSim_vs_SF}, with an increase of $SF$ values, both the BLEU and semantic similarity scores improve. This trend highlights enhanced robustness against interference, with the SRN-assisted SS-based SemCom system consistently outperforming the baseline across all $SF$ values. Accordingly, the effectiveness of the SRN in mitigating RFI impact is clearly confirmed. 

Of note, while the SRN-assisted system demonstrates robust performance, the E2E-trained DeepSC SemCom system exhibits a steeper performance gradient despite the increase in the number of transmitted symbols proportional to $SF$. The suboptimal performance of the online E2E training approach can be attributed to the high RFI, which corrupts the signal and degrades the quality of channel estimation during the training process. Additionally, it is worth noting that with an increase of \(SF\), the effective bandwidth used by the system also increases. This expansion in bandwidth proportionally increases the thermal noise power, as the total thermal noise power in a communication system is given by the product of noise spectral density (\(N_0\)) and bandwidth ($B$).

Fig. \ref{fig:signal_estimation} compares signal estimation capabilities. The proficiency of the SRN in closely approximating the original transmitted signal is evident in Fig. \ref{fig:signal_estimation}, despite high RFI and imperfect channel estimation. Conversely, the received and equalized signal, shown with a dashed line, significantly deviates from the original, further highlighting the need for and efficiency of the SRN in enhancing signal fidelity in environments containing high RFI.
\begin{figure}[t]
    \centering
    \includegraphics[width=0.96\columnwidth]{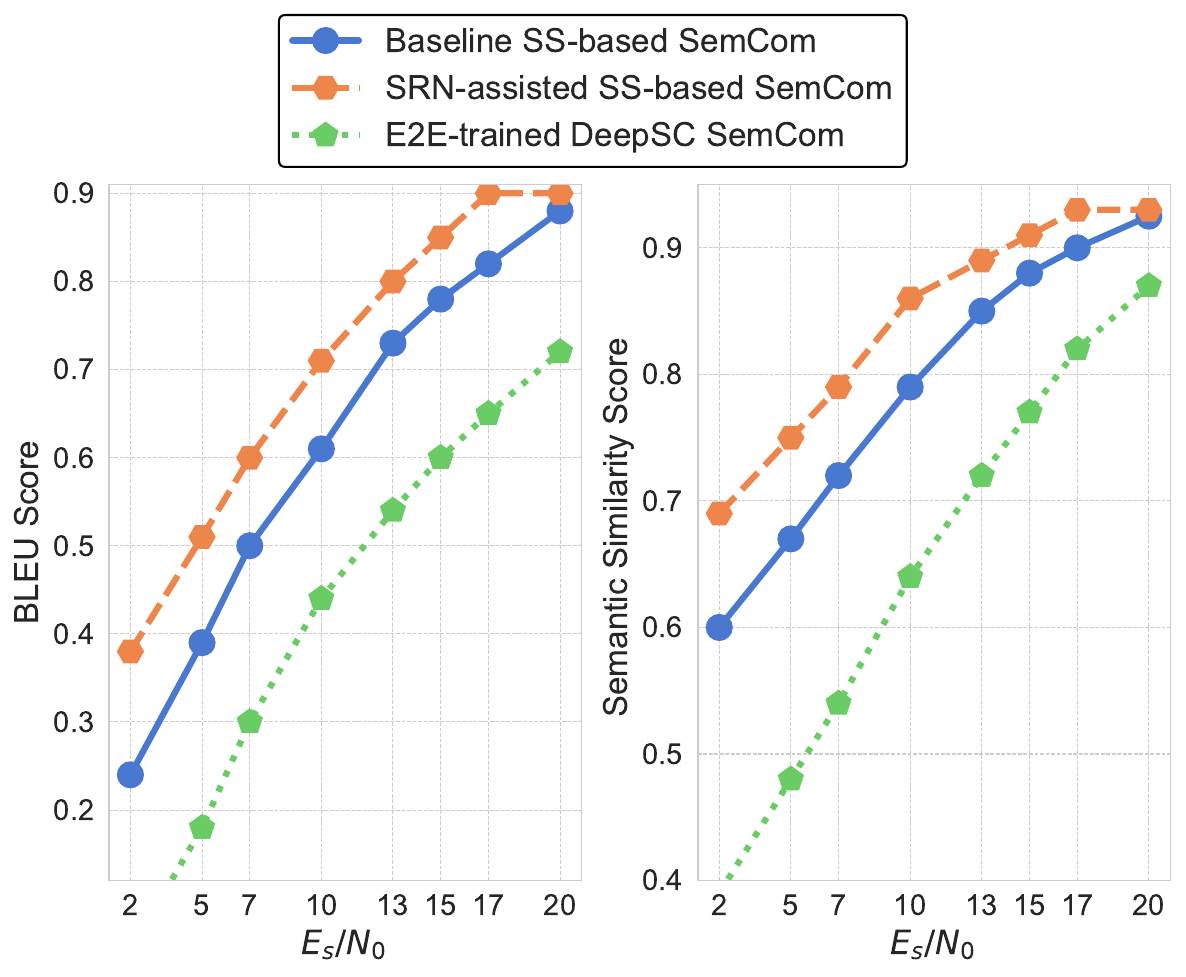}
    \caption{Performance comparison of baseline SS-based SemCom, SRN-assisted SS-based SemCom, and E2E-trained DeepSC SemCom systems across different \( E_s/N_0 \) values in the SU scenario, with $SF = 4$. }
    \label{fig:Performance_SNR}
\end{figure}
\subsection{Effect of SNR in SU Scenarios}

Similar to observations on the impact of the spreading factor (\( SF \)) in subsection \ref{subsection:SF_effect}, the effects of the SNR exhibit analogous trends in the performance of SemCom systems (see Fig. \ref{fig:Performance_SNR}). As can be seen in Fig. \ref{fig:Performance_SNR}, all systems exhibit improvements in both BLEU and semantic similarity scores as \( E_s/N_0 \) increases. Specifically, the SRN-assisted SS-based SemCom system consistently outperforms the baseline system, demonstrating a notable advantage of approximately 3 dB. Importantly, the performance advantage of the SRN-assisted system narrows with an increase in the SNR, eventually becoming negligible at very high SNR levels. This trend is expected, as the influence of RFI and noise naturally diminishes in scenarios with high SNR, reducing thus the need for the SRN's processing capabilities.

\subsection{Effect of SNR in MU Scenarios}
\begin{figure}[t]
    \centering
    \includegraphics[width=0.96\columnwidth]{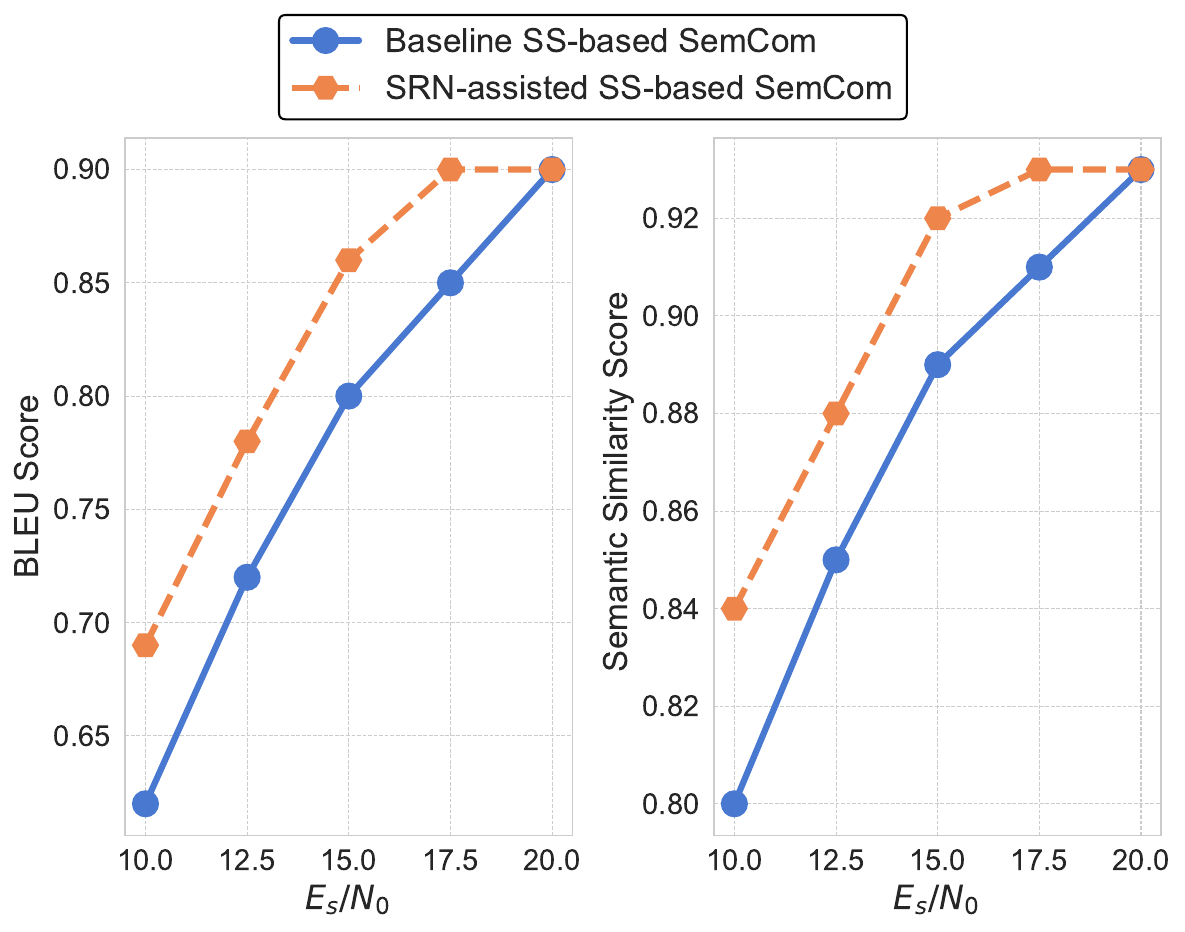}
    \caption{Performance comparison of baseline and SRN-assisted SS-based SemCom systems at different \( E_s/N_0 \) values in an MU scenario with six users (\( M = 6 \)), \( SF = 10 \), \( D = 0.25 \), and \( \mathbf{c}_i \cdot \mathbf{c}_j^H = 0 \) \(\forall i \neq j\) under Rayleigh fading and high RFI conditions.}
    \label{fig:MU_SemSim_vs_SNR}
\end{figure}
In the MU scenario, characterized by the use of orthogonal spreading codes, performance trends are similar to those observed in SU scenarios (see Fig. \ref{fig:MU_SemSim_vs_SNR}). This similarity in performance underlines the effectiveness of our proposed SRN-assisted SS-based SemCom system in MU scenarios. 

Importantly, the proposed system achieves robust performance without requiring joint or E2E online training of the transceivers, which significantly enhances its scalability and reduces the needed online training time (see Table \ref{tab:training_time_comparison}, where we have an average over 20 trials). Additionally, the complexity of DL-based SemCom systems may seem high because they contain several neural networks. However, in reality, it is comparable to traditional Turbo-coded bit-based communication systems, as demonstrated in \cite{deepsc}.

\begin{table}[t]
\centering
\caption{Average Online Training Time Comparison for One Transceiver with $SF = 8$ using Intel(R) Core(TM) i7-11800H CPU @ 2.30GHz.}
\label{tab:training_time_comparison}
\resizebox{1\columnwidth}{!}{
\renewcommand{\arraystretch}{1.05}
\begin{tabular}{|c|c|}
\hline
\textbf{Method} & \textbf{Average online training time (minutes)} \\
\hline
Baseline SS-based SemCom & No online training required \\
\hline
SRN-assisted SS-based SemCom & 3 \\
\hline
E2E-trained DeepSC SemCom & 35 \\
\hline
\end{tabular}}
\end{table}

\section{Conclusion}

In this study, we introduced a robust AI-assisted SS-based SemCom framework for environments affected by RFI. Testing results demonstrate marked improvements in both SU and MU scenarios, significantly reducing the need for extensive E2E or joint training. 

While we used generalized pseudo-random codes as a proof of concept, the findings highlight a pathway for further research. Future investigations could focus on applying specific coding strategies such as Gold codes, Walsh codes, or more advanced non-binary and polyphase sequences, which are likely to also offer improved robustness and interference mitigation in densely populated network environments.

\bibliographystyle{IEEEtran}
\bibliography{main} 

\end{document}